\documentclass{article}
\usepackage{spconfarXiv,amsmath,graphicx}
\usepackage{subfigure}
\usepackage{colortbl}
\usepackage{bm}
\usepackage{stfloats}
\usepackage{graphicx}  
\usepackage{epstopdf}
\usepackage{url}       
\usepackage{amsmath}   
\usepackage{amsfonts,amssymb}
\usepackage{cases}
\usepackage{algorithm}
\usepackage{multirow}
\usepackage{algorithmic}
\usepackage{microtype}
\usepackage{setspace}
 \usepackage{enumerate}
\usepackage{amsthm}
\usepackage[style=ieee, doi=false, isbn=false, url=false]{biblatex}
\newcommand{\ls}[1]
    {\dimen0=\fontdimen6\the\font
     \lineskip=#1\dimen0
     \advance\lineskip.5\fontdimen5\the\font
     \advance\lineskip-\dimen0
     \lineskiplimit=.9\lineskip
     \baselineskip=\lineskip
     \advance\baselineskip\dimen0
     \normallineskip\lineskip
     \normallineskiplimit\lineskiplimit
     \normalbaselineskip\baselineskip
     \ignorespaces
    }

\definecolor{purple}{RGB}{128,0,128}
\newcommand{\att}{\textcolor{black}}

\title{{Channel State Information-Free} Artificial Noise-Aided Location-Privacy Enhancement\vspace{-12pt}}

\name{Jianxiu Li and Urbashi Mitra\vspace{-15pt}\thanks{ This work has been funded in part by one or more of the following: NSF CCF-1817200, ARO W911NF1910269, DOE DE-SC0021417, Swedish Research Council 2018-04359, NSF CCF-2008927, NSF CCF-2200221, ONR 503400-78050, ONR N00014-15-1-2550, NSF CCF 2200221 and the USC Amazon Trust Center.} \thanks{Authors' emails: \texttt{\{{jianxiul,ubli}\}@usc.edu}}}
\address{Ming Hsieh Department of Electrical Engineering, Viterbi School of Engineering\\
University of Southern California, Los Angeles, CA, USA\vspace{-20pt}} 

\let\tilde\widetilde

\begin{document}
%
\maketitle
\begin{abstract}
\vspace{-5pt}
{In this paper, an artificial noise-aided strategy is presented for location-privacy {preservation}. A novel framework for the reduction of location-privacy leakage is introduced, where {\it structured artificial noise} is designed to degrade the structure of the illegitimate devices' channel, without the aid of channel state information {at the transmitter}. Then, based on the {\it location-privacy enhancement} framework, a transmit beamformer is proposed to efficiently inject the structured artificial noise. Furthermore, the securely {\it shared information} is {characterized to enable {the} legitimate devices to localize accurately}. Numerical results 
show {a $9$dB degradation of illegitimate devices' localization accuracy} {is achieved}, and validate the efficacy of {\textit{structured}} artificial noise {versus} {\textit{unstructured}} Gaussian noise.}
\end{abstract}

\vspace{-5pt}
\begin{keywords}
{Location-privacy, structured artificial noise, beamforming.}
\end{keywords}

\vspace{-10pt}
\section{Introduction}
\label{sec:intro}
\vspace{-8pt}
With {the} deployment of the Internet-of-Things, {the} location of a device is becoming an important commodity. To achieve accurate localization, millimeter-wave (mmWave) signaling has been {considered in} \cite{Shahmansoori,Zhou,Li} due to the massive available bandwidth \cite{Rappaport}. On the other hand, location-based
services as well as the proliferation of wearables necessitate location-privacy \cite{Zhang}.  However, most works on localization, {\it e.g.,} \cite{Shahmansoori,Zhou,Li}, {focus on} estimation accuracy without {considering location-privacy}. The location-privacy of user equipment {can be} significantly jeopardized by illegitimate devices that maliciously eavesdrop {on} the wireless channel. 

To combat eavesdropping, location-privacy preserving strategies {have predominantly been} studied at the application and network layers \cite{Yu, Tomasin, Schmitt}. To {the best of our knowledge}, the potential of physical-layer signals for protecting the location-privacy of the user equipment {has not been} fully investigated. {While} \cite{Checa, Tomasin2,Goel} exploit artificial interference {and} beamformer {design} for secrecy capacity improvement, these recent location-privacy preserving methods strongly rely on perfect knowledge of channel state information (CSI). \att{In particular, without the aid of CSI, the artificially injected Gaussian noise cannot be designed to lie in the {\it null space} of the channel, degrading the efficacy of \cite{Goel,Tomasin2}.} In the sequel, motivated by the analysis of the structure of the channel {in} \cite{Beygi, Elnakeeb, Li,Li2}, we propose a {structured artificial noise} (SAN)-aided {location-privacy enhancement} scheme to reduce location-privacy leakage (LPL) to illegitimate devices, while providing localization guarantees to authorized devices. In contrast to \cite{Goel, Checa,  Tomasin2}, our proposed design does not require CSI, which avoids extra channel estimation and reduces the overhead for resource-limited user equipment. The main contributions of this paper are:\vspace{-6pt}
\begin{enumerate}[1)]
\item {The proposal of} a general framework for SAN-aided location-privacy enhancement is without CSI, where the intrinsic structure of the channel is explicitly exploited to design the artificial noise; \vspace{-8pt}
\item Based on the proposed location-privacy enhancement framework, a beamforming strategy is designed to efficiently inject SAN, where LPL is mitigated, while authorized devices maintain localization accuracy with the securely {shared information} (a secret key);\vspace{-8pt}
\item Numerical comparisons {of} the localization accuracy of authorized devices show that the proposed method {results in} around $9$dB accuracy degradation for illegitimate devices, even when {illegitimate devices know} the structure of the designed artificial noise and line-of-sight (LOS) path exists. Furthermore, for low signal-to-noise ratios (SNR), the proposed scheme is more effective {for} location-privacy {preservation}, than the injection of unstructured Gaussian noise, with less shared information.
\end{enumerate}

\vspace{-18pt}
\section{System Model}\label{sec:signal}
\vspace{-10pt}
We consider a legitimate receiver (Bob) serving a user equipment (Alice), as shown in Fig. 1. The locations of Alice and Bob are denoted by $\bm p=[p_x,p_y]^{\mathrm{T}}\in \mathbb{R}^{2}$ and $\bm q=[q_x,q_y]^{\mathrm{T}}\in \mathbb{R}^{2}$, respectively. To acquire location-based services, Alice transmits pilot signals to Bob, while Bob estimates Alice's position based on the received signals and his location $\bm q$. We assume that the pilot signals are known {to} Bob. An unauthorized receiver (Eve) exists at location $\bm z=[z_x,z_y]^{\mathrm{T}}\in \mathbb{R}^{2}$, which is assumed to be close to Bob. {Eve} also knows the {same} pilot signals and her location $\bm z$. By eavesdropping {on} the channel to infer $\bm p$, Eve jeopardizes Alice's location-privacy.

\begin{figure}[t]
    \centering
    \includegraphics[width=0.49\textwidth]{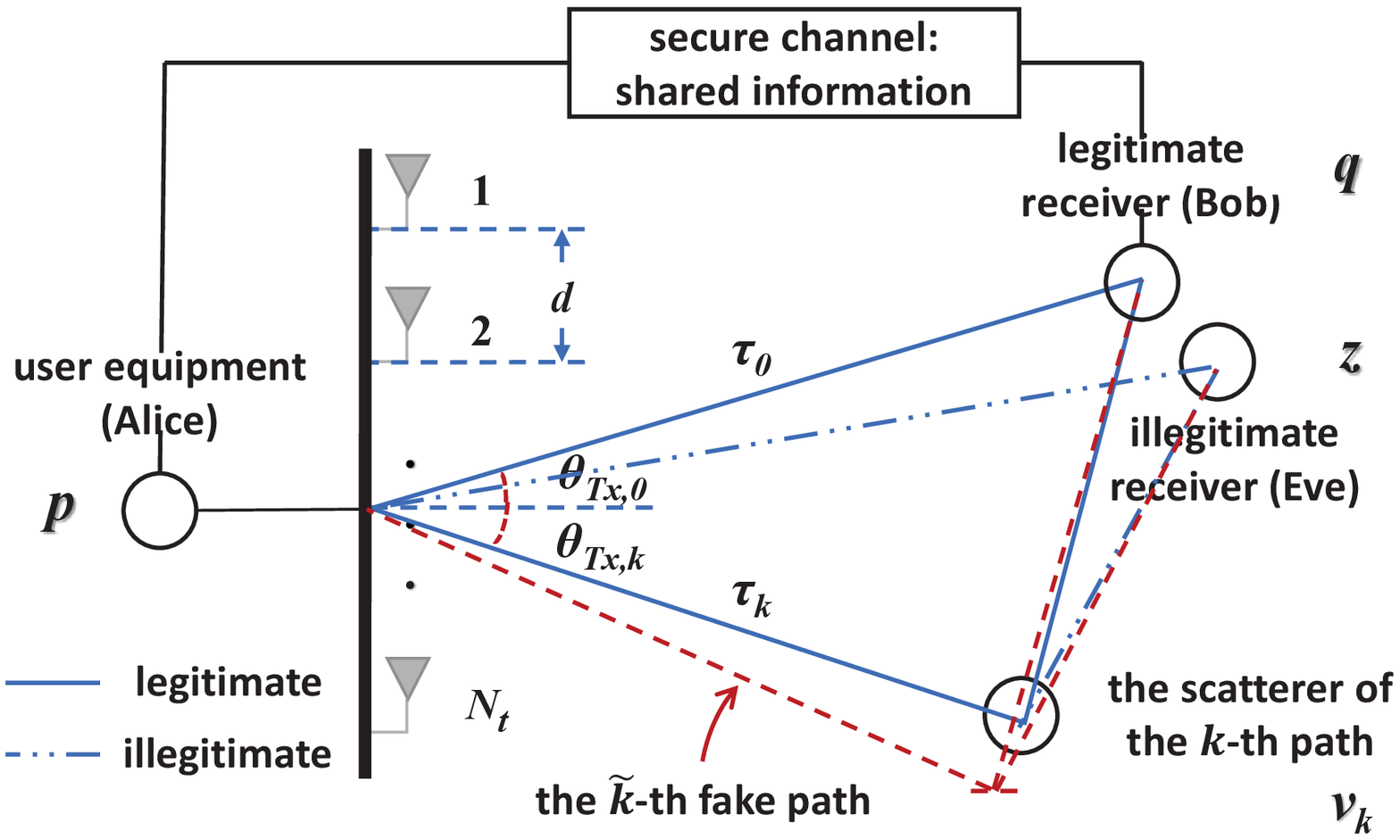}\vspace{-15pt} 
    \caption{{ System model.}}
    \label{systemmodel}\vspace{-15pt}
\end{figure}

Without of loss of generality, we adopt the mmWave multiple-input-single-output (MISO) orthogonal frequency-division multiplexing (OFDM) channel model of \cite{Fascista} for transmissions, where Alice is equipped with $N_t$ antennas, while both Bob and Eve have a single antenna\footnote{{The proposed location-privacy enhancement framework in Section \ref{sec:framework} can be directly extended to the multiple-input-multiple-output channel.}}. Assume that $K+1$ paths, {\it i.e.,} one LOS path and $K$ non-line-of-sight (NLOS) paths, exist in the MISO OFDM channel, and the scatterer of the $k$-th NLOS path is located at an unknown position $\bm v_k=[v_{k,x},v_{k,y}]^{\mathrm{T}}\in \mathbb{R}^{2}$, for $k=1,2,\cdots,K$. We transmit $G$ OFDM pilot signals via $N$ sub-carriers with carrier frequency $f_c$ and bandwidth $B$. It is assumed that a narrowband channel is employed, {\it i.e.,} $B\ll f_c$. Denoting by $x^{(g,n)}$ and $\bm f^{(g,n)}\in\mathbb{C}^{N_t\times 1}$ the $g$-th symbol transmitted over the $n$-th sub-carrier and the corresponding beamforming vector, respectively, we can express the $g$-th pilot signal over the $n$-th sub-carrier as
$\boldsymbol{s}^{(g,n)}\triangleq \bm f^{(g,n)}x^{(g,n)}\in\mathbb{C}^{N_t\times 1}$ and write the received signal ${y}^{(g,n)}$ as\vspace{-6pt}
\begin{equation}
{y}^{(g,n)}=\boldsymbol{h}^{(n)}  \boldsymbol{s}^{(g,n)}+{w}^{(g,n)},\label{rsignal}\vspace{-6pt}
\end{equation}
for $n=0,1,\cdots,N-1$ and $g=1,2,\cdots,G$, where ${w}^{(g,n)}\sim \mathcal{CN}({0},\sigma^2)$ is an independent, zero-mean, complex Gaussian noise with variance $\sigma^2$, and $\bm h^{(n)}\in\mathbb{C}^{1\times N_t}$ represents the $n$-th sub-carrier public channel vector. Denote by $c$, $d$, $T_s$ and $\bm{a}_L(f) \in \mathbb{C}^L$ the speed of light, the distance between antennas, the sampling period $T_s\triangleq\frac{1}{B}$, and the Fourier vector
\(
    \bm{a}_L(f) \triangleq  \left[1, e^{-j 2\pi f}, \dots, e^{-j 2\pi (L-1)f}\right]^{\mathrm{T}}
\), respectively. The public channel vector $\bm h^{(n)}$ is defined as\vspace{-6pt}
\begin{equation}
\boldsymbol{h}^{(n)}\triangleq\sum_{k=0}^{K}\gamma_k e^{\frac{-j 2\pi n\tau_k}{N T_{s}}}\boldsymbol{ \alpha}\left(\theta_{\mathrm{Tx},k}\right)^{\mathrm{H}},
\label{channelmatrix_subcarrier}\vspace{-6pt}
\end{equation}
where $k=0$ corresponds to the LOS path, $\gamma_k$ represents the complex channel coefficient of the $k$-th path, while the steering vector $\boldsymbol{\alpha}(\theta_{\mathrm{Tx}})$ is defined as  $\boldsymbol{\alpha}(\theta_{\mathrm{Tx}}) \triangleq \bm{a}_{N_t}\left(\frac{d \sin(\theta_{\mathrm{Tx}})}{\lambda_c} \right)$ with $\lambda_c\triangleq\frac{c}{f_c}$ being the wavelength. The operator $[\cdot]^{\mathrm{H}}$ represents the Hermitian transpose. According to the geometry, the location-relevant channel parameters of each path,  {\it i.e.,} time-of-arrival (TOA) $\tau_{k}$ and angle-of-departure (AOD) $\theta_{\mathrm{Tx}, k}$, are
$\tau_{k} =\frac{\left\|\boldsymbol{v}_0-\boldsymbol{v}_{k}\right\|_{2} +\left\|\boldsymbol{p}-\boldsymbol{v}_{k}\right\|_{2}} {c}$
and $\theta_{\mathrm{Tx}, k} =\arctan \left(\frac{v_{k,y}-p_{y}} {v_{k,x}-p_{x}}\right)
$, for $k=0,1,\cdots,K$, where $\bm v_0\triangleq \bm q$ (or $\bm v_0\triangleq \bm z$) holds for Bob (or Eve). In addition, we assume $\frac{\tau}{NT_s } \in(0,1]$ and ${\frac{d \sin \left(\theta_{\mathrm{Tx}}\right)}{\lambda_{c}} }\in(-\frac{1}{2},\frac{1}{2}]$ as {in} \cite{Li}.

Given the pilot signals, the location of Alice can be estimated using all the received signals. Hereafter, we denote by ${y}^{(g,n)}_{\text{Bob}}$ and ${y}^{(g,n)}_{\text{Eve}}$ the signals leveraged by Bob and Eve, respectively, which are defined according to \eqref{rsignal}, while the public channels for Bob and Eve, denoted as $\boldsymbol{h}^{(n)}_{\text{Bob}}$ and $\boldsymbol{h}^{(n)}_{\text{Eve}}$, are modelled according to \eqref{channelmatrix_subcarrier}, {as a function of} their locations. Note that {the} CSI is assumed to be unavailable {to Alice}. {We aim} to reduce {the} LPL to Eve with the designed artificial noise, providing localization guarantees to Bob with the shared information transmitted over a secure channel.

\vspace{-10pt}
\section{Structured Artificial Noise-Aided Location-Privacy Enhancement }\label{sec:framework}
\vspace{-8pt}
Since it is assumed that CSI is unknown, to preserve location-privacy, we create {a SNR advantage} for Bob versus Eve via the injection of artificial noise tailored to the intrinsic structure of the channel. In this section, we present a general framework for location-privacy enhancement with {SAN}.

According to the analysis of the atomic norm based localization {in} \cite{Li}, high localization accuracy relies on super-resolution channel estimation; to ensure the quality of the estimate, the TOAs and AODs of the $K + 1$ paths need to be sufficiently separated, respectively. Inspired by \cite{Beygi, Elnakeeb, Li,Li2}, we propose SAN to degrade the structure of the channel for location-privacy preservation, where the minimal separation for TOAs and AODs, {\it i.e.,} $\Delta_{\min}(\frac{\tau}{NT_s})$ and $\Delta_{\min}\left(\frac{d \sin \left(\theta_{\mathrm{Tx}}\right)}{\lambda_c}\right)$, with $\Delta_{\min}(\kappa)\triangleq\min_{i\neq j}\min(|\kappa_i-\kappa_j|,1-|\kappa_i-\kappa_j|)$, is reduced, respectively\footnote{{For the atomic norm based localization method in \cite{Li}, the conditions $\Delta_{\min}(\frac{\tau}{NT_s})\geq\frac{1}{\lfloor\frac{N-1}{8}\rfloor}$ and $\Delta_{\min}\left(\frac{d \sin \left(\theta_{\mathrm{Tx}}\right)}{\lambda_c}\right)\geq\frac{1}{\lfloor\frac{N_t-1}{4}\rfloor}$ are desired.}}. To be more precise, in this framework, we virtually add $\tilde{K}$ {\it fake paths} to the mmWave MISO OFDM channel, as shown in Fig. \ref{systemmodel}. Denote by $\tilde{\gamma}_{\tilde{k}}$, $\tilde{\tau}_{\tilde{k}}$, and $\tilde{\theta}_{\mathrm{Tx},\tilde{k}}$ the artificial channel coefficient, TOA, and AOD of the $\tilde{k}$-th {\textbf{fake}} path, respectively. For the $g$-th pilot signal transmitted over the $n$-th sub-carrier, the injected SAN $\xi^{(g,n)}$ is designed as,\vspace{-7pt}
\begin{equation}
{\xi}^{(g,n)}\triangleq\tilde{\bm h}^{(n)}\boldsymbol{s}^{(g,n)},
\label{san}\vspace{-7pt}
\end{equation}
where \vspace{-10pt}
\begin{equation}
    \tilde{\bm h}^{(n)}\triangleq\sum_{\tilde{k}=1}^{\tilde{K}}\tilde{\gamma}_{\tilde{k}} e^{\frac{-j 2\pi n\tilde{\tau}_{\tilde{k}}}{N T_{s}}}\boldsymbol{ \alpha}\left(\tilde{\theta}_{\mathrm{Tx},\tilde{k}}\right)^{\mathrm{H}}\in\mathbb{C}^{1\times N_t}.
    \label{fakechannel}\vspace{-7pt}
\end{equation}
With the injection of SAN defined in \eqref{san}, the received signal {used} by Eve is\vspace{-7pt}
\begin{equation}
\begin{aligned}
{y}^{(g,n)}_{\text{Eve}}&={\bm h}^{(n)}_{\text{Eve}}  \boldsymbol{s}^{(g,n)}+ \xi^{(g,n)}+{w}_{\text{Eve}}^{(g,n)}\\
&=\left({\bm h}^{(n)}_{\text{Eve}}+\tilde{\bm h}^{(n)}\right)\boldsymbol{s}^{(g,n)}+{w}_{\text{Eve}}^{(g,n)}.
\end{aligned}\label{rsignalnEve}\vspace{-7pt}
\end{equation}
where ${w}_{\text{Eve}}^{(g,n)}\sim \mathcal{CN}({0},\sigma_{\text{Eve}}^2)$. Note that, given $\tilde{K}\geq 1$, the minimal separation for TOAs and AODs can be reduced according to the definition of $\Delta_{\min}(\cdot)$ {provided above}. 
On the other hand, to alleviate the distortion caused by SAN for Bob, we assume Alice transmits the {shared information}\footnote{{The shared information is determined by the design for the injection of SAN; one specific design will be provided in Section \ref{sec:method}. }} to Bob over a secure channel that is inaccessible by Eve. Then, the received signal leveraged by Bob for localization is given by\vspace{-7pt}
\begin{equation}
{y}^{(g,n)}_{\text{Bob}}={\bm h}^{(n)}_{\text{Bob}}  \tilde{\boldsymbol{s}}^{(g,n)}+{w}_{\text{Bob}}^{(g,n)},\label{rsignalnBob}\vspace{-2pt}
\end{equation}
where ${w}_{\text{Bob}}^{(g,n)}\sim \mathcal{CN}({0},\sigma_{\text{Bob}}^2)$. The signal $\tilde{\boldsymbol{s}}^{(g,n)}\in\mathbb{C}^{N_t\times 1}$ is determined by ${\boldsymbol{s}}^{(g,n)}$ and the shared information, which is known to Bob to maintain localization accuracy. 

In contrast to Bob {who can} {remove} {the SAN with the shared information}, the injected SAN distorts the structure of the channel and thus degrades the eavesdropping ability of Eve. We note that the proposed location-privacy enhancement framework does not rely on CSI or any specific estimation method\footnote{{The proposed framework can be extended for the design of secure communications.}}. In the next section, an efficient strategy for the injection of SAN will be shown, where the virtually added fake paths are further elaborated {upon}, while the specific design of $\tilde{\gamma}_{\tilde{k}}$, $\tilde{\tau}_{\tilde{k}}$, $\tilde{\theta}_{\mathrm{Tx},\tilde{k}}$, $\tilde{\boldsymbol{s}}^{(g,n)}$ and {the shared information} are provided. 

\vspace{-10pt}
\section{Beamforming Design for the Injection of structured Artificial Noise}\label{sec:method}
\vspace{-9pt}
Based on the location-privacy preserving framework in Section \ref{sec:framework}, SAN is injected via adding fake paths. However, it is {impractical} to {create} extra scatterers for the fake paths. In the sequel, we design a transmit beamforming strategy to efficiently reduce the LPL to Eve. 
\vspace{-10pt}
\subsection{Alice's Beamformer}\vspace{-7pt}
Let $\delta_\tau$ and $\delta_{\theta_{\text{TX}}}$ represent two parameters used for beamforming design. To enhance the location-privacy, Alice still {employs} the mmWave MISO OFDM signaling according to Section \ref{sec:signal}, but designs {her transmitter} beamformer $\tilde{\bm f}^{(g,n)}$ as,\vspace{-3pt}
\begin{equation}
\begin{aligned}
    \tilde{\bm f}^{(g,n)} \triangleq \left(\bm I + e^{-\frac{2\pi n \delta_{\tau}}{NT_s}}\operatorname{diag}\left(\bm \alpha\left(\delta_{\theta_\text{Tx}}\right) \right)\right){\bm f}^{(g,n)},
\end{aligned}\label{fakebeamformer}\vspace{-3pt}
\end{equation}
for the $g$-th pilot signal transmitted over the $n$-th sub-carrier, where $\bm I\in\mathbb{R}^{N_t\times N_t}$ is the identity matrix and the operator $\operatorname{diag}(\bm a)$ represents a diagonal matrix whose diagonal elements are given by vector $\bm a$.
\vspace{-10pt}
\subsection{Bob's Localization}\label{Bobloc}\vspace{-7pt}
Through the public channel ${\bm h}^{(n)}_{\text{Bob}}$, Bob receives\vspace{-3pt}
\begin{equation}
{y}^{(g,n)}_{\text{Bob}}={\bm h}^{(n)}_{\text{Bob}}  \tilde{\boldsymbol{f}}^{(g,n)}x^{(g,n)}+{w}_{\text{Bob}}^{(g,n)},\label{rsignalbBob}\vspace{-3pt}
\end{equation}
for $n=0,1,\cdots,N-1$ and $g=1,2,\cdots,G$. Assume Bob has the knowledge of the structure of Alice's beamformer {in \eqref{fakebeamformer}, {\it i.e.,} the construction of $\tilde{\bm f}^{(g,n)}$ based on ${\bm f}^{(g,n)}$.} By leveraging the transmissions over the secure channel, Bob also knows {$\bm\delta\triangleq[\delta_\tau,\delta_{\theta_{\text{TX}}}]^\mathrm{T}\in\mathbb{R}^2$ that is shared information}. Given the pilot signal $\bm s^{(g,n)}$, he can construct the signal $\tilde{\bm s}^{(g,n)}$ as\vspace{-5pt}
\begin{equation}
\begin{aligned}
    \tilde{\bm s}^{(g,n)}&\triangleq \tilde{\boldsymbol{f}}^{(g,n)}x^{(g,n)}\\
    &=\left(\bm I + e^{-\frac{2\pi n \delta_{\tau}}{NT_s}}\operatorname{diag}\left(\bm \alpha\left(\delta_{\theta_\text{Tx}}\right) \right)\right){\bm s}^{(g,n)},
\end{aligned}\vspace{-3pt}
\end{equation}  
which simplifies \eqref{rsignalbBob} into \eqref{rsignalnBob}, {{\it i.e.,} ${y}^{(g,n)}_{\text{Bob}}={\bm h}^{(n)}_{\text{Bob}}  \tilde{\boldsymbol{s}}^{(g,n)}+{w}_{\text{Bob}}^{(g,n)}$}. As a result, with the securely shared information, the degradation of Bob's channel structure is negligible, {which aids his} localization accuracy. {Note that the amount of shared information $\bm\delta$ does not increase with respect to the number of the received signal, {\it i.e.,} $N$ or $G$, which is determined by the quantization and coding strategies. For simplicity, Bob is assumed to receive the perfect $\bm\delta$ through the secure channel.}
\subsection{Eve's Localization}\vspace{-7pt}
Assume that the shared information $\bm\delta$ is recycled at a certain rate such that Eve can not decipher it. Then, the following received signal has to be {used} if Eve attempts to estimate Alice's position, \vspace{-7pt}
\begin{equation}
\begin{aligned}
{y}^{(g,n)}_{\text{Eve}}&={\bm h}^{(n)}_{\text{Eve}}  \tilde{\boldsymbol{f}}^{(g,n)}x^{(g,n)}+{w}_{\text{Eve}}^{(g,n)}\\
&=\left({\bm h}^{(n)}_{\text{Eve}}+\tilde{\bm h}^{(n)}\right)  {\boldsymbol{s}}^{(g,n)}+{w}_{\text{Eve}}^{(g,n)}\\
&={\bm h}^{(n)}_{\text{Eve}}{\boldsymbol{s}}^{(g,n)}+{ \xi}^{(n)} +{w}_{\text{Eve}}^{(g,n)},\label{rsignalbEve}
\end{aligned}\vspace{-5pt}
\end{equation}
where ${ \xi}^{(n)}$ and $\tilde{\bm h}^{(n)}$ are defined in \eqref{san} and \eqref{fakechannel}, respectively, with\vspace{-5pt} $\tilde{K}=K$, $\tilde{\gamma}_{{k}}={\gamma}_{{k}}$,
\begin{equation}
    \tilde\tau_k=\tau_k+\delta_\tau,\label{tildetau}\vspace{-5pt}
\end{equation}
and \vspace{-3pt}
\begin{equation}
    {\tilde\theta_{\text{Tx},k}=\arcsin(\sin(\theta_{\text{Tx},k})+\sin(\delta_{\theta_{\text{Tx}}})).}\label{tildethetatx}\vspace{-15pt}
\end{equation} 
\vspace{-10pt}
\subsection{Minimal Separation and Location-Privacy Leakage}\label{subsec:mslpl}\vspace{-7pt}
{Given that the values of $\delta_\tau$ and $\delta_{\theta_{\text{Tx}}}$ are small enough, the minimal separation for TOAs and that for AODs are $\left|\frac{\delta_\tau}{NT_s}\right|$ and $\left|\frac{d\sin(\delta_{\theta_\text{Tx}})}{\lambda_c}\right|$, respectively, according to \eqref{tildetau}, \eqref{tildethetatx} and the definition of $\Delta_{\min}(\cdot)$.}   Coinciding with the goal of the framework proposed in Section \ref{sec:framework}, the minimal separation for TOAs and AODs can be efficiently decreased with the design of parameters $\delta_\tau$ and $\delta_{\theta_{\text{TX}}}$\footnote{The optimal choice of these design parameters depends on the locations as well as the pilot signals, wihch is beyond the scope of this paper, but small values are suggested without CSI. } and thus
{the structure of the channel is effectively degraded}. 


{Using the signals received over the SAN-distorted channel as defined in \eqref{rsignalbEve}, it is more difficult for Eve to accurately estimate Alice's position without the access to shared information}, even in the present of {the} LOS path, which will be further verified in Section \ref{sec:sim}. {We define the LPL as}\footnote{\att{$\text{LPL}\leq 0$ means that Eve cannot infer Alice's position more accurately than Bob; $\text{LPL}> 0$ means that Eve can achieve higher localization accuracy.}},\vspace{-5pt}
\begin{equation}
    \text{LPL}\triangleq\frac{\text{RMSE}_\text{Bob}-\text{RMSE}_\text{Eve}}{\text{RMSE}_\text{Bob}}\label{LPL},\vspace{-5pt}
\end{equation}
where $\text{RMSE}_\text{Bob}$ and $\text{RMSE}_\text{Eve}$ represent the root-mean-square error (RMSE) of localization achieved by Bob and Eve.

\vspace{-10pt}
\section{Numerical Results}\label{sec:sim}
\vspace{-9pt}
To show the degradation of Eve's eavesdropping ability with the proposed method, in this section, we evaluate the Cram\'{e}r-Rao lower bound (CRLB) of the estimation of Alice's position for Bob and Eve, respectively. With respect to LPL, we consider the worst case where the structure of Alice's beamformer in \eqref{fakebeamformer} is exposed to Eve. In contrast to Bob who exploits ${y}^{(g,n)}_{\text{Bob}}$ and $\tilde{\bm  s}^{(g,n)}_{\text{Bob}}$ to estimate $\left\{\bm p,\left\{\bm v_k\right\},\left\{ \gamma_k\right\}\right\}$, the goal of Eve is to infer $\left\{\bm p,\left\{\bm v_k\right\},\left\{ \gamma_k\right\},\delta_\tau,\delta_{\theta_{\text{TX}}}\right\}$ using ${y}^{(g,n)}_{\text{Eve}}$ and ${\bm  s}^{(g,n)}_{\text{Eve}}$. Due to lack of space, the details regarding the CRLB are omitted. We refer {the reader} to \cite{Fascista} for {related} derivations. In all of the numerical results, the system parameters $B$, $f_c$, $c$,  $N_t$, $N$, $G$, $K$, and $d$ are set to $15$ MHz, $60$ GHz, $300$ m/us, $16$, $16$, $16$, $2$, and $\frac{\lambda_c}{2}$, respectively.  The free-space path loss model \cite{Goldsmith} is used to {determine} channel coefficients in the simulation, while the pilot signals are random, complex values uniformly generated on the unit circle. The scatterers of the two NLOS paths are located at $[8.89\text{ m}, -6.05 \text{ m}]^{\mathrm{T}}$ and $[7.45 \text{ m}, 8.54 \text{ m}]^{\mathrm{T}}$, respectively, while Alice is at $[3 \text{ m},0 \text{ m}]^{\mathrm{T}}$. To make a fair comparison, Bob and Eve are placed at the same location, {\it i.e.,} $[10 \text{ m},5 \text{ m}]^{\mathrm{T}}$, and the same received signal is used for the simulation. The SNR is defined as $10\log_{10}\frac{\sum^{G}_{g=1}\sum^{N-1}_{n=0}|{\bm h}^{(n)}_{\text{Bob}}  \tilde{\boldsymbol{s}}^{(g,n)}|^{2}_{}}{NG\sigma^2}$\footnote{The SNR is also equal to $10\log_{10}\frac{\sum^{G}_{g=1}\sum^{N-1}_{n=0}|\left({\bm h}^{(n)}_{\text{Eve}}+\tilde{\bm h}^{(n)}\right) {\boldsymbol{s}}^{(g,n)}|^{2}_{}}{NG\sigma^2}$ for a fair comparison.}, where the operator $|\cdot|$ represents the the magnitude of a complex value. Unless otherwise stated, the design parameters $\delta_\tau$ and $\delta_{\theta_{\text{TX}}}$ are set to $-\frac{\pi}{61}$ and $10^{-8}$, respectively.

\begin{figure}[t]
\centering\vspace{-6pt}
\includegraphics[scale=0.48]{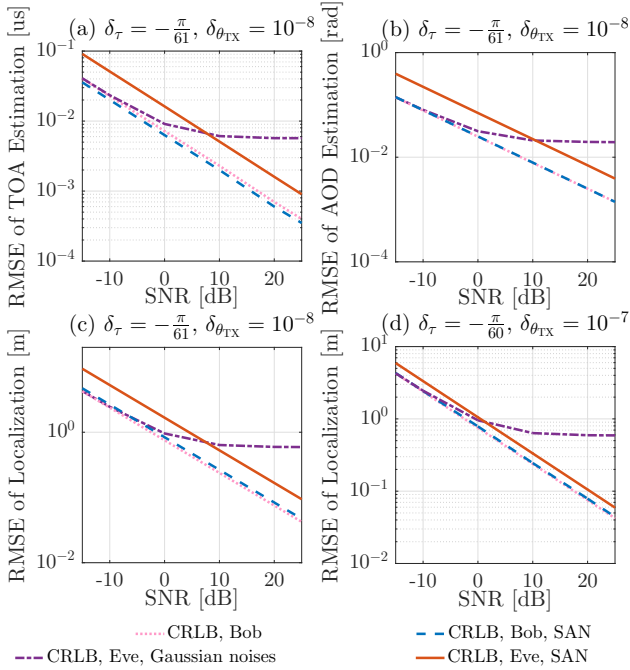}\vspace{-13pt}
\caption{\att{The RMSE of channel estimation and  localization.}}%
\label{numericalresults}\vspace{-10pt}
\end{figure}
\begin{figure}[t]
\centering
\includegraphics[scale=0.55]{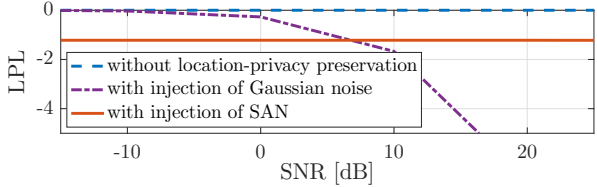}\vspace{-15pt}
\caption{{Comparisons of LPL.}}%
\label{numericalresultslpl}\vspace{-15pt}
\end{figure}

The RMSEs of TOA estimation and AOD estimation are shown in Figs. \ref{numericalresults} (a) and (b), respectively, where the CRLB for Bob's estimation is compared with that for Eve's estimation. As observed in Figs. \ref{numericalresults} (a) and (b), our proposed scheme contributes to more than $9$dB degradation with respect to the TOA and AOD estimation, by virtue of the fact that SAN effectively distorts the structure of the channel; the negative effect is negligible for Bob who has the securely shared information \att{to remove SAN, with the comparison to the case without the injection of any artificial noise}. Due to the strong degradation of the quality of channel estimates, {a larger} CRLB for Eve's localization is achieved as seen in Figs. \ref{numericalresults} (c). With respect to the localization accuracy, there is around {a $9$dB advantage} for Bob versus Eve using our proposed scheme. {In addition, the LPL defined in \eqref{LPL} is reduced to around $-1.22$ according to Fig. \ref{numericalresultslpl}, in contrast to the case without location-privacy preservation ($\text{LPL}=0$)}, which {suggests} location-privacy enhancement. To show the influence of the choices of the design parameters for Alice's beamformer, the CRLBs {for} localization are plotted in Fig. \ref{numericalresults} (d), with increased values of $|\delta_{\tau}|$ and $|\delta_{\theta_{\text{TX}}}|$. As compared with the results in Fig. \ref{numericalresults} (c), the gap between Bob and Eve's performance is 
reduced due to the {increase of the minimal separation based on the analysis in Section \ref{subsec:mslpl}}. The optimal choices of $\delta_\tau$ and $\delta_{\theta_{\text{TX}}}$ will be investigated in future work.

To validate the efficacy of SAN, CRLBs for channel estimation and localization with the injection of additional Gaussian noise are also provided in Figs. \ref{numericalresults} and \ref{numericalresultslpl} as comparisons. For fair comparison, the variance of the artificially added Gaussian noise $\zeta^{(g,n)}\sim \mathcal{CN}({0},\varsigma^2)$ is set to a constant value for all SNRs, {\it i.e.,} $\varsigma^2\triangleq\frac{\sum^{G}_{g=1}\sum^{N-1}_{n=0}|\tilde{\bm 
h}^{(n)} {\boldsymbol{s}}^{(g,n)}|^{2}_{}}{NG}$, while the SNR is still defined as previously stated. As observed in Figs. \ref{numericalresults} and \ref{numericalresultslpl}, the injection of the extra Gaussian noise is ineffective to preserve location-privacy at low SNRs since its constant variance is relatively small. In contrast, through the degradation of the channel structure, the proposed artificial noise strongly enhances  location-privacy. As SNR increases, the injection of the additional Gaussian noise can further degrade the Eve's performance due to the constant variance. However, only the worst case of our scheme {has been shown here}, where the structure of the designed beamformer is assumed to be accessible by Eve. Furthermore, to remove the artificially injected Gaussian noise for Bob \att{without CSI}, the amount of securely shared information, {{\it i.e.}, $\zeta^{(g,n)}$}, increases with respect to $NG$, in contrast to our scheme only leveraging {two parameters} $\delta_{\tau}$ and $\delta_{\theta_{\text{TX}}}$, for a given quantization and coding strategy.

\vspace{-10pt}
\section{Conclusions}\label{sec:con}
\vspace{-9pt}
{A location-privacy enhancement strategy was investigated with the injection of artificial noise. A novel CSI-free location-privacy enhancement framework was proposed, where the structure of the channel was exploited for the design of SAN. To effectively preserve location-privacy based on the proposed framework, a transmit beamformer was 
designed to reduce the LPL to unauthorized devices, while authorized devices can maintain localization accuracy using the securely shared information. The proposed method strongly degraded the eavesdropping ability of unauthorized devices. With respect to the CRLB for localization, there was around $9$dB degradation in contrast to authorized devices with the shared information. Furthermore, the efficacy of SAN was numerically verified with the comparison to unstructured Gaussian noise.}

%


\vfill
\pagebreak



%
\renewcommand*{\bibfont}{\small}
\printbibliography


\end{document}